\documentclass[preprint2]{aastex}
\usepackage{graphicx}
\usepackage{pdflscape}
\usepackage{float}
\usepackage{morefloats}

\shorttitle{The anomalous extinction of the ionizing cluster in
NGC\,3603} \shortauthors{Pang et al.}

\begin{document}

\title{The grey extinction of the ionizing cluster in NGC\,3603 from ultraviolet to optical wavelengths}


\author{Xiaoying Pang\altaffilmark{1,2,3,4}, Anna Pasquali\altaffilmark{3}, Eva K.\ Grebel\altaffilmark{3}}

\affil{$^{1}$Shanghai Institute of Technology, 100 Haiquan Road,
Fengxian district, Shanghai 201418, China}

\affil{$^{2}$National Astronomical Observatories, Chinese Academy of
Sciences, 20A Datun Road, Chaoyang District, 100012 Beijing, China}
\affil{$^{3}$Astronomisches Rechen-Institut, Zentrum f\"ur
Astronomie der
          Universit\"at Heidelberg, M\"onchhofstr.\ 12--14, 69120
              Heidelberg, Germany}
\affil{$^{4}$ Key Lab for Astrophysics, 100 Guilin Road, Shanghai
200234, China}
 \email{xypang@bao.ac.cn}




\begin{abstract}
We use photometry in the {\em F220W}, {\em F250W}, {\em F330W}, {\em
F435W} filters
 from the High Resolution Channel of the Advanced Camera for Surveys
 and photometry in the {\em F555W}, {\em F675W}, and {\em F814W} filters from the Wide Field and
Planetary Camera 2 aboard the Hubble Space Telescope to derive
individual stellar reddenings and extinctions for stars in the
HD\,97950 cluster in the giant H\,{\sc ii} region NGC\,3603. The
mean line-of-sight reddening for about a hundred main-sequence
member stars inside the cluster is
$E(F435W-F555W)=1.33\pm0.12$\,mag. After correcting for foreground
reddening, the total to selective extinction ratio is
$R_{F555W}=3.75\pm0.87$ in the cluster.
 Within the standard deviation associated with $E(\rm \lambda-F555W)/E(F435W-F555W)$
 in each filter, the cluster extinction curve at ultraviolet wavelengths tends to be greyer than the average Galactic extinction
laws from Cardelli et al.\ (1989) and Fitzpatrick et al.\ (1999). It
is closer to the extinction law derived by Calzetti et al.\ (2000)
for starburst galaxies, where the 0.2175\,$\rm \mu m$ bump is
absent. This indicates an anomalous extinction in the HD\,97950
cluster, which may due to the clumpy dust distribution within the
cluster, and the size of dust grains being larger than the average
Galactic ISM.
\end{abstract}




\section{Introduction}
The dependence of dust extinction on wavelength is described by the so-called extinction curve.
 In most regions of the Milky Way, the extinction curve rises like a power law from infrared (IR) to optical wavelengths,
 then shows a prominent bump at 0.2175\,$\rm \mu m$, and may steeply rise in the far ultraviolet (UV) (Figure 1).
 Possible contributors to the UV bump are interstellar graphite
grains (Mathis et al.\ 1977), polycyclic aromatic hydrocarbon (PAH) molecules
(Draine 1989) molecules, or MgO particles with a thin O$^{2-}$
coating (Maclean et al.\ 1982).

A typical way to determine the interstellar extinction is the
standard ``pair method'', which matches the spectral features of
reddened main-sequence (MS) stars with identical unreddened MS
standard stars (Massa et al.\ 1983). The extinction is derived for
each star in terms of
 $A_{\rm \lambda}/A_{V}$ or $E(\lambda-V)/E(B-V)$ as a function of inverse wavelength
(Cardelli et al. 1988, 1989; Fitzpatrick 1999). Recently Fitzpatrick
\& Massa (2005) introduced a new technique that derives the
interstellar extinction curve from the UV to the IR by modeling the observed
spectral energy distribution of reddened early-type stars.

 The Galactic extinction curve has been extensively studied in a large number of papers
(e.g., Seaton 1979; Savage et al. 1985; Cardelli et al.\ 1989, 1988;
Fitzpatrick 1999; Valencic et al.\ 2004; Fitzpatrick \& Massa 2007,
2009), which have shown it to vary considerably among different
lines of sight.
The individual extinction curves obtained for various lines of sight are averaged to obtain the mean Galactic extinction curve,
 which has an average total to selective extinction ratio of $R_{V}=3.1$, and whose dependence on wavelength is
parameterized with analytical functions.
While the extinction curve at optical wavelengths does not seem to change much among different lines of sight,
 large variations exist in the UV range (Massa et al.\ 1983; Fitzpatrick 1999).
The large rms dispersion 
associated with the different average Galactic extinction curves
makes them almost indistinguishable from each other
(Seaton 1979; Savage et al.\ 1985; Cardelli et al.\ 1989;
Fitzpatrick 1999; Valencic et al.\ 2004; Fitzpatrick \& Massa 2007).
 We should bear in mind that any mean extinction curve
is biased by the sample of lines of sight from which it was derived,
 i.e., different lines of sight probe different conditions of the interstellar medium (ISM).

However, the mean Galactic extinction curve does not work for other
galaxies, especially not for starburst galaxies, where the star
formation rate is a few tens of $\rm M_\odot yr^{-1}$, and the total to
selective extinction ratio $R_{V}$ is much larger ($\sim4.05$;
Calzetti et al.\ 1994, 1997, 2000) than in the diffuse Galactic ISM
($R_{V}=3.1$). The extinction curve of starburst galaxies is grey
and lacks a bump at 0.2175\,$\rm \mu m$ (Figure 1). The value of
$R_{V}$ is usually treated as an indicator for the grain size
(Cardelli et al. 1989). The higher than normal value of $R_{V}$ and
the absence of a UV bump may imply different processing histories of
dust grains in the ISM of starburst galaxies, which favor formation
of large grains.

According to Calzetti et al.\ (1994), young massive stars are
associated with dustier regions than the older and less massive
stars. The large optical depth associated with young massive stars
preserves the large dust grains, which absorb less UV light than
smaller grains and thus may explain the absence of the UV bump in
the extinction curve of starburst regions/galaxies (Pierini et al.\
2005; Panuzzo et al.\ 2007). Furthermore, the strong UV radiation
and stellar winds of massive stars are likely to destroy dust grains
and change the properties of the local ISM. The stellar feedback
from young massive stars produces an inhomogeneous, clumpy
distribution of dust within star-forming regions, which cannot be
represented with the uniform dust screen model commonly used for the
Milky Way. A clumpy distribution of dust is effective in redirecting
the scattered photons along the line of sight (Calzetti et al.
1994). In contrast, in the diffuse Galactic ISM the distant dust
forms a uniform foreground extinction screen, through which 
absorption and the scattering remove flux from the line of
sight. Therefore the extinction curve of starburst galaxies is
greyer than the average Galactic extinction curve.

Unfortunately, most of the starburst galaxies are too distant to
study their ISM with high angular resolution data. However, in the
Milky Way, the environment of an H\,{\sc ii} region hosting a
massive young cluster may resemble that of starburst galaxies. A
systematic investigation of the reddened stars within an H\,{\sc ii}
region can probe the effect of the presence of massive stars on dust
grains, and extend our knowledge of the ISM in starburst regions
from parsec to kiloparsec scales.

 There is an ideally suited
Galactic H\,{\sc ii} region hosting a starburst cluster of which
high-resolution and multi-band data are available. This is
NGC\,3603, a giant H\,{\sc ii} region containing the central
ionizing cluster HD\,97950 with a mass of $\sim\rm 10^4\,M_\odot$
(Harayama et al.\ 2008). A larger than normal total to selective
extinction $R_V=4.3$ was found among stars in NGC\,3603 that are
brighter than $V=17$\,mag and located at distances larger than
0.5\,pc from the HD\,97950 cluster (Pandey et al.\ 2000). Sung \&
Bessell (2004) derived a smaller $R_V$ value, $3.55\pm0.12$,
 from stars with $V\le16.5$\,mag that are located at projected cluster-centric
radii of $r\ge0.4$\,pc. Values of $R_V$ larger than 3.1 were also
found in a number of other young Galactic star clusters in H\,{\sc
ii} regions,
 e.g., NGC\,6530 (Fernandes et al.\ 2012);
M\,16, M\,17, and NGC\,6357 (Chini \& Kr\"{u}gel 1983; Chini \&
Wargau 1990). A larger value of $R_V$ can be due to significant
changes in dust properties, such as evaporation of small grains by
the radiation of hot stars (Draine 2009) and growth of large grains
(Cardelli et al. 1989; Hirashita 2012).

The strong radiation field of the OB stars in the HD\,97950 cluster (Drissen et al.\ 1995; Melena et al.\ 2008) ionizes and
 sweeps away the ISM, and generates a cavity around the cluster where the gas reddening is the lowest (Pang et al.\ 2011).
A shell structure was found at 1.2\,pc west of the cluster (Figure 2, Clayton 1986; Pang et al.\ 2011), which
 expands with a velocity of 55\,$\rm km\,s^{-1}$. The engine of the expanding shell may be the stellar
winds and the radiation pressure of the massive stars in the cluster. Such energy input into
the ISM can change the dust grain properties significantly.
Weingartner \& Draine (2001) found that when grains larger than
$\rm 0.1\,\mu m$ are exposed to the radiation of OB stars, they are subjected to the
photodesorption force, applied by a radiation field that
 is anisotropic because of a spatially uneven distribution of stars.
Under this force, the large grains decouple from the gas and move
away. Furthermore, the large grains may be destroyed or reduced to
75\% of their initial radius by a shock speed as low as 40\,$\rm
km\,s^{-1}$, which can be generated by supernovae (Seab \& Shull
1983) or by the radiation pressure and the stellar winds of OB stars
in a cluster. The large grains are destroyed or sputter into small
grains and are pushed away.
 This agrees with the observation of Lebouteiller et al. (2007)
that the emission of very small grains whose size is smaller
than $\rm 0.02\,\mu m$ (Wood et al. 2008) increases from the cluster
center outwards and peaks at cluster-centric distances of 1 -- 1.5\,pc.

Spectroscopy is only available for 26 early-type stars in NGC\,3603
(Moffat 1983; Drissen et al.\ 1995; Melena et al.\ 2008). Using
$UBV$ photometry, Pandey et al.\ (2000) derived stellar reddenings
and extinctions for 51 probable member stars, whose memberships were 
inferred via photometric methods and which are mainly located in
the outskirts of the cluster HD\,97950 in NGC\,3603. Sung \& Bessell
(2004) extended the number of stars with stellar reddening $E(B-V)$
in NGC\,3603 to a few hundreds. No membership probability estimation
is available for these stars.

In our work, we obtain reddenings and extinctions for about a
hundred MS member stars in HD\,97950. The membership of these stars
was determined via relative proper motions in Pang et al.(2013). We
also derive the average extinction curve from UV to optical
wavelengths for the member stars within the HD\,97950 cluster, and
compare it to the extinction laws in the Milky Way and in starburst
galaxies.

In Section 2 we will introduce the observations and data reduction.
We will discuss the individual stellar reddening of cluster stars in
Section 3. We discuss the total to selective extinction ratio within
the HD\,97950 cluster in Section 4. A discussion of the inferred dust
growth history and the extinction curve is presented in Section 5. Finally, we
summarize our results in Section 6.

\section{Observations and Data reduction}

The HD\,97950 cluster and its immediate surroundings in NGC\,3603 were observed with
 the Hubble Space Telescope (HST). The UV data were taken with the High Resolution
Channel (HRC) of the Advanced Camera for Surveys (ACS) in 2005 (GO
10602, PI: Jes\'{u}s Ma\'{i}z Apell\'{a}niz) through the {\em
F220W}, {\em F250W}, {\em F330W}, and {\em F435W} filters. The HRC
is characterized  by a spatial resolution of $0.03''$ pixel$^{-1}$
and a field of view of 29$''$ $\times$ 25$''$. The optical
observations were carried out with the Wide Field and Planetary
Camera 2 (WFPC2) in two epochs: 1997 (GO 6763, PI: Laurent Drissen)
and 2007 (GO 11193, PI: Wolfgang Brandner) through the {\em F555W},
{\em F675W}, and {\em F814W} filters. The Planetary Camera (PC) chip
was centered on the cluster ($0.045''$\,pixel$^{-1}$,
$40''\times40''$) for both programs. The wavelength range of these
filters is shown in Figure 1. The {\em F220W} filter covers the UV
bump in the extinction curve, which can effectively distinguish a
normal from a grey extinction curve.
Pang et al.\ (2013) reduced the two-epoch WFPC2 data and
identified more than 400 member stars on the PC chip via relative proper motions.
Of these member stars, 142 are in common between the HRC and PC
images and thus
 have UV and optical photometry available (see Table 1).

The ACS/HRC data were reduced using {\it Dolphot} (Dolphin 2005), a stellar
photometry package that was adapted from {\it HSTphot} (Dolphin 2000) to handle ACS images.
 We used the ACS module in {\it Dolphot} for the {\em F220W}, {\em F250W}, {\em F330W} and
{\em F435W} data. All the data consist of four
dithered images. The exposure time for each single exposure image
(*.flt) is 357\,s in the {\em F220W}, 44\,s in the {\em F250W}, 10\,s in the {\em F330W} and 2\,s in the {\em F435W} filter.
We masked bad and saturated pixels in the
``flt'' images with the routine ACSMASK.
The offsets between dithered images were calculated with a python
script. The photometry was produced by the routine {\it Dolphot}, with
the specific parameters suggested for ACS data (Dolphin 2006). {\it Dolphot} performs point spread function
(PSF) photometry with the precomputed PSF
of each filter.

The WFPC2 data were reduced with {\it HSTphot}
(Dolphin 2000), which uses a similar reduction procedure as {\it Dolphot}.
 A detailed description of the data reduction of the optical WFPC2 data is provided in Pang et al.\ (2013).


\section{Individual stellar reddenings in the HD\,97950 cluster}

It is crucial to know which stars are cluster members in order to
estimate the reddening of a given star cluster (e.g., Yadav \& Sagar
2001). Among the HD\,97950 cluster member stars determined from
relative proper motions (Pang et al.\ 2013, Table 2), there are five
MS stars located in the cluster with projected distances of
$r<0.7$\,pc from the center, for which there are also spectral types
available from Table 3 of Melena et al.\ (2008). We indicate the
locations of these five stars (red circles) in the cluster in Figure
2. The photometry of these five MS stars is presented in Table 2.

In order to obtain the reddening of these five stars, we derive
their intrinsic colors from an empirical zero age main sequence
(ZAMS) according to their spectral types. The MS stars within the
HD\,97950 cluster were found to be consistent with a single burst
population with an age of 1 Myr (e.g., Kudryavtseva et al.\ 2012;
Pang et al.\ 2013). Hence a ZAMS provides a good approximation of
their properties in color-magnitude space. We use the empirical ZAMS
from Sparke \& Gallagher (S\&G, 2007, their Table 1.4), which is
based on a large number of nearby MS stars with reliably determined
photometry, distances, and well-defined spectral types. The S\&G
ZAMS only provides {\em UBV} photometry in the Johnson filter
system.

On the other hand, the Marigo et al.\ (2008) theoretical isochrones
provide magnitudes not only in the standard Johnson filter system,
but also in the ACS/HRC and WFPC2 systems, which are appropriate for
the present data sets. These isochrones extend to 90\,M$_\odot$,
which is massive enough to cover the most massive member stars used
in this paper.
 Hence we derive the corresponding {\em F220W}, {\em F250W}, {\em F330W}, {\em
F435W}, {\em F555W}, {\em F675W}, and {\em F814W} magnitudes for
each spectral type from a Marigo et al.\ (2008) 1\,Myr-old isochrone
of solar metallicity. This is easily doable since the Marigo et al.\
isochrones of a given age and metallicity use exactly the same
luminosity, temperature, and mass steps for each photometric system.
We can now plot two-color diagrams corresponding to the traditional
{\em UBV} two-color diagram, and obtain the appropriate reddening
values for our five stars (Figure 3) by subtracting the intrinsic
colors (blue points) from the observed colors (red points).

Repeating this procedure for different color combinations allows us
to calculate the color excesses $E(F220W-F434W)$, $E(F250W-F435W)$,
$E(F330W-F435W)$, $E(F435W-F555W)$, $E(F555W-F675W)$, and
$E(F555W-F814W)$, and the extinctions $A_{F220W}$, $A_{F250W}$,
$A_{F330W}$, $A_{F435W}$, $A_{F555W}$, $A_{F675W}$, and $A_{F814W}$
by comparing the intrinsic colors and magnitudes obtained from the
ZAMS using the observed apparent magnitudes and a distance modulus
of 14.2\,mag from Sung \& Bessell (2004).

The  ({\em F555W}-{\em F814W}) vs.\ {\em F555W} color-magnitude diagram (Figure 4) shows
the member stars of the HD 97950 cluster based on the membership
determinations by Pang et al. (2013).  Black dots denote stars
detected both in the optical and in the UV data.  Open circles show
those stars with optical detections only.  Member stars with UV and
optical photometry delineate the massive MS of the HD 97950 cluster.
The fainter stars with optical detections only are either MS stars
or pre-main-sequence (PMS) stars.

Note the wide ``turn-on'' transition region where PMS stars are
moving towards the MS in the color range ({\em F555W}-{\em
F814W})=$1.8-2.6$\,mag.  MS stars located in this overall region are
contaminated with PMS stars. Some of the brightest PMS stars are
also detected in our UV data (and accordingly are marked with black
dots).
 Since our procedure to compute individual reddening values relies on MS
stars, we exclude PMS stars from our analysis. Moreover, calculating
reddenings for PMS stars is fraught with uncertainties due to the
unknown effects caused by their circumstellar disk or dust shell
absorbing the UV radiation, the inclination of such disks,
accretion, etc (Baraffe et al. 2009).

 In order to obtain a clean sample of MS stars, we
consider only stars bluer than ({\em F555W}-{\em F814W})=1.8\,mag
 to be MS stars. The lowest
stellar mass of these objects corresponds to
 $\rm \sim4\,M_\odot$.
 Altogether, we select 111 MS member stars to study the reddening and extinction curve within the cluster.

We use the mean reddening vector $E(F330W-F435W)/E(F435W-F555W) =
0.984 \pm 0.014$ (the error is the standard deviation) of the five
MS stars selected from Melena et al.\ (2008) to project the 111
member MS stars onto the 1\,Myr-old Marigo et al.\ (2008) isochrone
in color-color space and derive their individual $E(F330W-F435W)$
and $E(F435W-F555W)$ (Figure 5). The uncertainty of the color excess
comes from the uncertainty in the reddening vector and the
photometry. The
 reddening vector contributes $\sim0.05$\,mag uncertainty to the value of
$E(F330W-F435W)$, $\sim0.01$\,mag to $E(F435W-F555W)$. The
individual color excesses and extinctions of the member MS stars are
listed in Table 3. Table 4 presents the corresponding mean values
for the whole sample.

Within the limitations of our photometry, we cannot estimate the
amount of multiplicity among the member stars. However, the optical
color of a star hotter than B7 ($\rm\sim4\,M_\odot$) will not be
significantly affected by an equal-mass or less massive companion.
The mean reddening inside the cluster,
$E(F435W-F555W)=1.33\pm0.12$\,mag (Table 4), agrees well with the
result of Sung \& Bessell (2004, $E(B-V)=1.25-1.4$\,mag for stars
within 0.7\,pc from the center), who derived the reddening by ZAMS
fitting to the CMD.
 We plot the $E(F435W-F555W)$ of the member MS stars as a function of
their cluster-centric distances in panel a of Figure 6. The error
bars indicate the uncertainty of the $E(F435W-F555W)$ values.
 There is an apparent
trend of increasing $E(F435W-F555W)$ values (red dots) with
cluster-centric radius, implying that differential reddening even
exists within the cluster (of $r\le0.7$\,pc), just as was found
earlier by Sung \& Bessell (2004; Figure 5 in their paper).

\section{Total to selective extinction ratio}

The mean ratio of total to selective extinction inside the cluster
is $R_{F555W}=3.54\pm0.63$ (see Table 4) computed from the
extinction $A_{F555W}$ and the color excess $E(F435W-F555W)$.
 The $R_{F555W}$ value of majority of the cluster member stars
  exceeds 3.1 (Table 3, see Figure 7).

We adopt the $E(B-V)$ value of 1.1\,mag obtained by Pandey et al.\
(2000) as an upper limit for the foreground reddening towards the
HD\,97950 cluster. After we correct for foreground reddening
(assuming Fitzpatrick 1999 with $R_V$=3.1), the mean $R_{F555W}$
increases to $R_{F555W}=3.75\pm0.87$. 
In Section 5.3, we have shown that the spectral type uncertainty in our selected five MS
member stars (see Table 2) does not significantly affect the final
$R_V$ value.



 To investigate the spatial dependence of dust properties in
HD\,97950, we plot $R_{F555W}$ and $A_{F555W}$ as a function of
cluster-centric distance in panels b and c of Figure 6, both of
which show significant scatter. We computed the average values of
$R_{F555W}$ and $A_{F555W}$ within concentric annuli, respectively,
with an initial radius of 0.1\,pc increasing to 0.5\,pc in steps of
0.1\,pc. The mean values are plotted as red dots in panels
b and c of Figure 6. The mean $R_{F555W}$ and $A_{F555W}$ values remain roughly
constant within the uncertainties, showing a possible slight
increase in the outermost
 annulus.

 The dispersion of $R_{F555W}$, $\sigma_{R_{F555W}}$, is larger in
the innermost region of the cluster than in the outer part (panel
c). The large variation of $A_{F555W}$ and $R_{F555W}$ inside the
cluster may imply that the column density of dust within the cluster
HD\,97950 and possibly the size of dust grains vary along different
lines of sight, resulting in a clumpy dust distribution across the
cluster. This could be the result of stellar feedback from massive
stars, which is destroying dust grains and modifying their
distribution.

\section{Discussion}

\subsection{Inferred dust growth history for the HD 97950 region}

 A high value of $R_V$ ($>3.1$) likely indicates that the size of
dust grains in the HD\,97950 cluster is larger than the mean grain
size in the diffuse ISM. The growth of dust grains, which may have
occurred via accretion or coagulation (Hirashita 2012), took place
in the parental molecular cloud of the cluster, where the dust
grains can accrete mantles of molecules (e.g., ice) or elements
(e.g., C, N, and O) from the gas phase (Cardelli et al. 1989).
Accretion of volatile ice mantles on the grains increases the
sticking coefficients in grain-grain collisions (Ossenkopf 1993),
hence enhancing the chance of coagulation, i.e., the adherence of
dust particles driven by their mutual interaction. Though a higher
cloud density promotes coagulation (Tielens 1989), the latter
depends also on the particles' collision velocity, which is a
function of mass, elastic properties, and surface energy of the dust
material (Ossenkopf 1993). Cardelli et al. (1989) computed the
integrated extinction cross sections in several dense Galactic
regions ($R_V>3.1$), and found that their extinction cross section
is smaller than that of the diffuse ISM. Since accretion increases
the total extinction cross section while coagulation either
increases or decreases it, they suggested that the size growth of
grains is dominated by coagulation, which is confirmed by a recent
study of Campeggio et al. (2007).

Whittet et al. (2001) found that when dust grows via coagulation,
$R_V$ increases on average with $A_V$ across the whole extinction
range, while mantle growth (accretion) requires a threshold
extinction below which $R_V$ is independent of $A_V$.
 In the cluster HD\,97950, the trend that $R_{F555W}$ increases almost linearly with $A_{F555W}$ (panel d in Figure 6)
 implies that large dust grains might have formed via coagulation in the parental
 cloud of this cluster.


\subsection{Extinction Curve within HD\,97950}

We have obtained extinctions in the {\em F220W}, {\em F250W}, {\em
F330W}, {\em F435W}, {\em F555W}, {\em F675W}, and {\em F814W}
filters for the cluster MS member stars (Table 3) in Section 3. We
adopt the foreground reddening from Pandey et al.\ (2000), with an
assumption of a Fitzpatrick (1999) extinction law for the foreground
($R_V=3.1$). In Figure 8, we compare the mean extinction curves
(corrected for foreground reddening) $E({\rm
\lambda}-F555W)/E(F435W-F555W)$ (left panel) and
$A_\lambda/A_{F555W}$ (right panel) for the 111 MS members (black
line) and the five MS stars with spectroscopy (red line) from which
we obtain the reddening vector of the cluster in Section 3.
 These two extinction curves agree quite well with each other,
 confirming the consistency of our method.

 In Figure 9, we plot the mean extinction curves
 for the MS members (black line) already corrected
 for the foreground extinction, and compare it with
 Cardelli et al.'s ($R_V=3.1$, 1989), Fitzpatrick's ($R_V=3.1$, 1999),
 and Calzetti et al.'s ($R_V=4.05$, 2000) extinction laws (red dashed lines).
We normalize the extinction curves to the visual color excess
$E(F435W-F555W)$ (upper panels) and extinction $A_\lambda/A_{F555W}$
(lower panels). As the $R_V$ of the HD\,97950 cluster is larger than
the mean Galactic value 3.1, we also plot the resulting extinction
laws of Fitzpatrick (1999), Cardelli et al. (1989) and Calzetti et
al. (2000) using the $R_V$ value of the cluster, 3.75, for
comparison (red solid lines). Calzetti et al.'s (2000) extinction
law is independent of $R_V$ when it is expressed as $E({\rm
\lambda}-F555W)/E(F435W-F555W)$. We indicate the uncertainties of Fitzpatrick's
(1999) and Calzetti et al.'s (2000) extinction laws with error
bars. Uncertainties are not provided in the study of Cardelli et al. (1989).

At optical wavelengths (filter passbands {\em F814W} to {\em F435W}) all 
three extinction law prescriptions are essentially indistinguishable 
from our data. 
In the mid-UV, within the standard deviation associated with $E({\rm
\lambda}-F555W)/E(F435W-F555W)$ (error bars), the extinction curve
(black line) we obtained for the HD\,97950 cluster (upper panels)
tends to be greyer than the Galactic extinction laws with $R_V=3.75$
(red dotted lines) and $R_V=3.1$ (red dashed lines, Cardelli et al.
1989; Fitzpatrick 1999). The cluster extinction curve
is closer to the extinction law of Calzetti et al. (2000),
especially in the {\em F220W} ($MUV$) pass bands (black box). However, we cannot
distinguish the cluster extinction curve normalized to the visual
extinction $A_\lambda/A_{F555W}$, from the laws of Fitzpatrick
(1999), Cardelli et al. (1989) and Calzetti et al. (2000) with the
same $R_V=3.75$ as the cluster (lower panels). Since a possible age
spread, unrecognized binaries, and errors in assigning spectral
types all may contribute to systematic errors in the absolute
magnitude measurements while colors are less affected, plotting
$E({\rm \lambda}-F555W)/E(F435W-F555W)$ is more sensitive for detecting
potential differences between extinction curves.

We have assumed that for the foreground extinction the standard
$R_V$ value holds, i.e., $R_{V_{fore}} = 3.1$. In order to explore
how a different ratio of total to selective extinction would affect
the resulting cluster extinction curve, we test other values.
Choosing a smaller $R_{V_{fore}}$ of 2.6 makes the cluster
extinction curve greyer, whereas it stays almost unchanged when
increasing $R_{V_{fore}}$ to 3.6.

Both the extinction laws of Cardelli et al. (1989) and Fitzpatrick
(1999) for $R_V$ values of 3.1 and 3.75 show a prominent UV bump at
0.2175\,$\rm \mu m$ (Figure 1 \& 9). However, the mean extinction
curve of the HD\,97950 cluster is similar to the grey extinction law
of starburst galaxies (Calzetti et al.\ 2000), which is
characterized by the absence of the 0.2175\,$\rm \mu m$ bump in the
mid-UV ({\em F220W}, Figure 1 \& 9). 
   To investigate the strength of the UV bump, we derive
    $E(F220W-F555W)/E(F435W-F555W)$ from the extinction laws of
    Fitzpatrick (1999) and Cardelli et al. (1989) computed with the individual $R_{F555W}$ values of
 all MS member stars (Table 3). The $E(F220W-F555W)/E(F435W-F555W)$
 values of the member stars are all
 smaller than those derived from Fitzpatrick's (1999, red line) and Cardell et
 al.'s (1989, black line) law, except for five stars (Figure 10). This points to a
 greyer extinction in the HD\,97950 cluster than the Galactic average
 one. 

Studies of Pierini et al.\
(2005) and Panuzzo et al.\ (2007) both found that the strength of
the UV bump decreases when dust is mostly composed of large grains
that formed in a very dense molecular cloud and absorb less UV
radiation with respect to smaller grains (Hirashita 2012).
As shown by Witt \& Gordon (2000) in the case of the Milky Way, a
clumpy dust distribution can lead to the greyness of the extinction
curve and the presence of a reduced UV bump.
 Our measurement of the color excess and extinction of stars
in the HD\,97950 cluster indicates that the dust distribution within
this cluster is likely clumpy (see Section 4).

\subsection{Individual Extinction Curve of member stars}

 As a consistency check, we plot in Figure 11
 the individual extinction curves of the five MS spectroscopic member
 stars, from whose spectral types we could derive independent
 color excesses and absolute extinctions. We compare the
 spectral types determined by Melena et al.\ (2008) and those by other
 studies (Moffat 1983, Drissen et al.\ 1995), and find a typical
 difference of two spectral sub-classes. Therefore, we assume that the
 uncertainty on the spectral type of these five stars to be two sub-classes, which corresponds to an uncertainty in the range of 
 $\sim0.002-0.02$ on the $E({\rm
 \lambda}-F555W)/E(F435W-F555W)$ (the size of the filled circles in Figure 11) of these five
stars.

The $R_V$ values of these stars span the range 2.48--4.06, and
 their visual extinctions range from 3.29 to 5.0\,mag.
Their extinction curves (in red) are compared with the
extinction laws of Cardelli et al.\ (1989), Fitzpatrick (1999) and
Calzetti et al.\ (2000) computed for their original $R_V=3.1$  and
4.05 (blue dashed lines),  for the stars' measured $R_V$ as
indicated in the upper panels (blue dotted lines), and the
extinction laws of Cardelli et al. \ (1989) and Fitzpatrick's  
(1999) extinction laws with $R_V$ values from 2.0 to
6.0 (grey shaded regions), which is the range of $R_V$ values for the stars
found when deriving Galactic extinction laws in Fitzpatrick's and Cardelli
et al's papers. The typical feature of these 
extinction laws is the existence of the bump at 0.2175\,$\rm \mu
m$ for all $R_V$ values.

 As shown in Figure 11, the extinction curves of these five stars with
 known spectral types appear
greyer than the extinction law of Fitzpatrick (1999) for the observed
$R_V$ values (2.0--6.0), and greyer than Cardelli et al.'s (1989)
law when calculated with the stars' $R_V$ values. Though Cardelli et
al.\ (1989) law with $R_V\sim5.0$ can reproduce the individual
extinction value for the values in the wavelength of  {\em F220W}, its values at the
wavelengths of 
{\em F250W} and {\em F330W} filters are below the observed values of the
five MS stars, resembling the feature of an UV bump. The absence of an
UV bump in the individual extinction curve closely mimicks 
the extinction law of Calzetti et al.\ (2000) independently of $R_V$.
The same can be seen when comparing the extinction laws that were
normalized to the visual extinction $A_{F555W}$. These results point
to the clumpiness of the dust distribution as the main cause for the
greyness of the stars' extinction curves.

\section{Summary}

We use the {\em F220W}, {\em F250W}, {\em F330W}, {\em F435W}, {\em
F555W}, {\em F675W}, and {\em F814W} photometry obtained with HST
HRC/ACS and WFPC2 to estimate individual stellar reddenings and
extinctions for stars in the HD\,97950 cluster. The main results of
our analysis are:

We derive stellar reddenings for 111 individual MS member stars
inside the HD\,97950 cluster. The mean line-of-sight reddening
inside the cluster is $E(F435W-F555W)=1.33\pm0.12$\,mag (with
$R_{F555W}=3.54\pm0.63$, without foreground reddening correction).
 Using only the value that is corrected for foreground extinction, the total to selective extinction ratio is
 $R_{F555W}=3.75\pm0.87$. The large dispersion of $R_{F555W}$ indicates a likely clumpy distribution of
 dust within the cluster.

 Within the standard deviation associated with $E({\rm \lambda}-F555W)/E(F435W-F555W)$ in each filter,
the extinction curve of the HD\,97950 cluster in the filter {\em
  F220W} corrected for foreground reddening
 tends to be greyer than the average Galactic extinction laws of Cardelli et al.\ (1989)
and Fitzpatrick (1999) with both $R_V=3.75$ and $R_V=3.1$. The
extinction curve is closer to the extinction law for starburst
galaxies (Calzetti et al.\ 2000), showing a flatter extinction trend
at the UV wavelengths. At longer 
wavelengths the cluster extinction curve agrees with all three
extinction laws.
 This may indicate an anomalous extinction law
in the HD\,97950 cluster, and that the dust in the HD\,97950 cluster
has similar properties to the dust in starburst galaxies. The
absence of the UV bump at 0.2175\,$\rm \mu m$ for the cluster and
starburst galaxies may be attributed to the clumpy dust
distribution, and, in addition, to dust grains that grew larger than
those in the average Galactic ISM because of the high density of the
cluster parental molecular cloud (Pierini et al.\ 2005; Panuzzo et
al.\ 2007).


\acknowledgments
 We thank Matthias Frank for his Python script to derive the offsets between ACS images.
 We are also grateful to the referee for his/her in-depth comments,
 which improved this paper.
This work is funded by National Natural Science
Foundation of China, No: 11503015, and the fund of Shanghai education
committee, No: 1021ZK151009027-ZZyy15104. 
XYP acknowledges travel support by
Sonderforschungsbereich 881 ``The Milky Way System'' of the German
Research Foundation (subproject B5). 
XYP is a member of the
Silk Road Project Team in National Astronomical Observatories of
China (NAOC, \texttt{http://silkroad.bao.ac.cn}), and acknowledges
the technical support from this team.

 \clearpage

\onecolumn


\clearpage

\begin{figure}[]
\includegraphics[width=11cm]{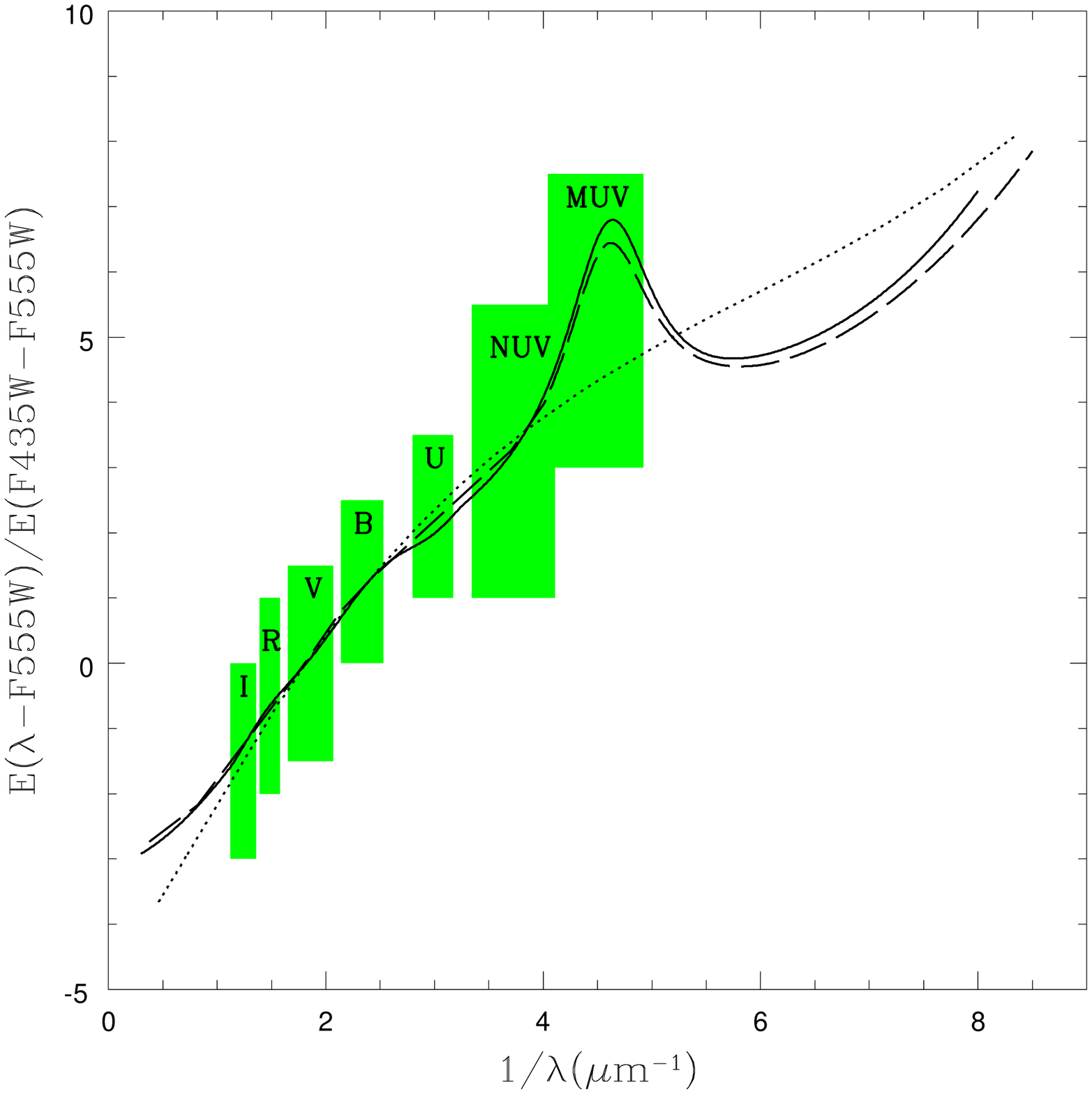}
\caption[] {Average Galactic extinction curves ($R_V$ = 3.1) of
Cardelli et al. (1989, solid line) and Fitzpatrick (1999, dashed
line). The dotted line is the mean extinction curve for 39 starburst
galaxies (Calzetti et al. 2000) with $R_V$ = 4.05. The UV bump in
the Galactic extinction curves is observed around 4.6\,$\rm \mu
m^{-1}$. The shaded areas indicate the wavelength coverage of each
filter. For convenience, we name the {\em F220W}, {\em F250W}, {\em
F330W}, {\em F435W}, {\em F555W}, {\em F675W}, and {\em F814W}
filters $MUV$, $NUV$, $U$, $B$, $V$, $R$, and $I$, respectively.
\label{Fig.5}}
\end{figure}

\begin{figure}[]
\includegraphics[width=\textwidth]{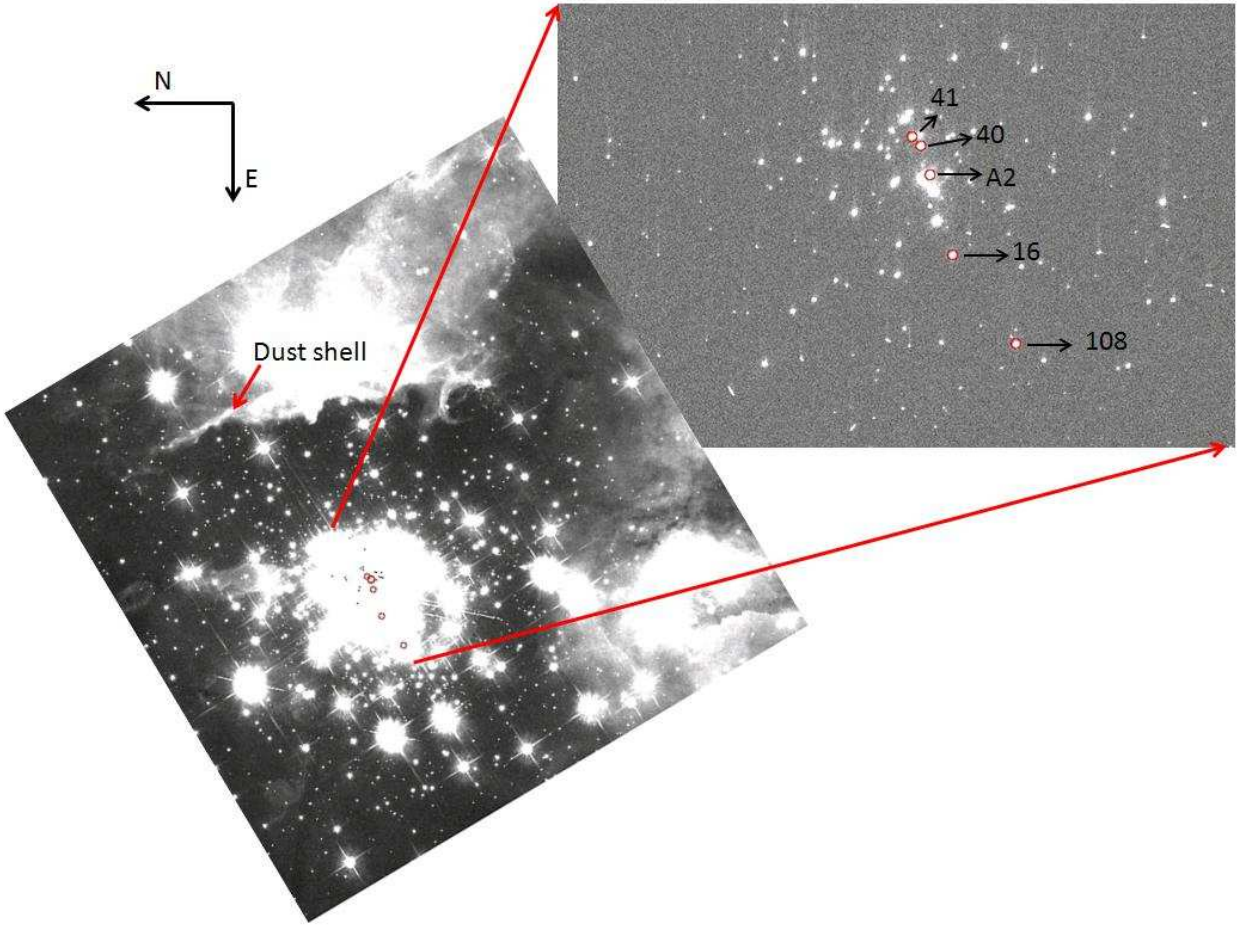}
\caption{Image of the cluster HD97950 in the {\em F555W} filter taken by
the HST WFC3 UVIS (bottom left, 162$''\times$162$''$) and WFPC2 PC
chip (upper right, 20$''\times$20$''$). The WFPC2 PC image is a
zoomed-in version of the cluster center which is saturated in the
WFC3 image. The five MS stars with spectral types available from
Melena et al. (2008) are indicated with red circles and their
designations given by Melena et al. (2008). The dust shell, which is
about 1.2\,pc away from the center, is indicated by an arrow.
 \label{Fig.10}}
\end{figure}

\begin{figure}[]
\includegraphics[width=11cm]{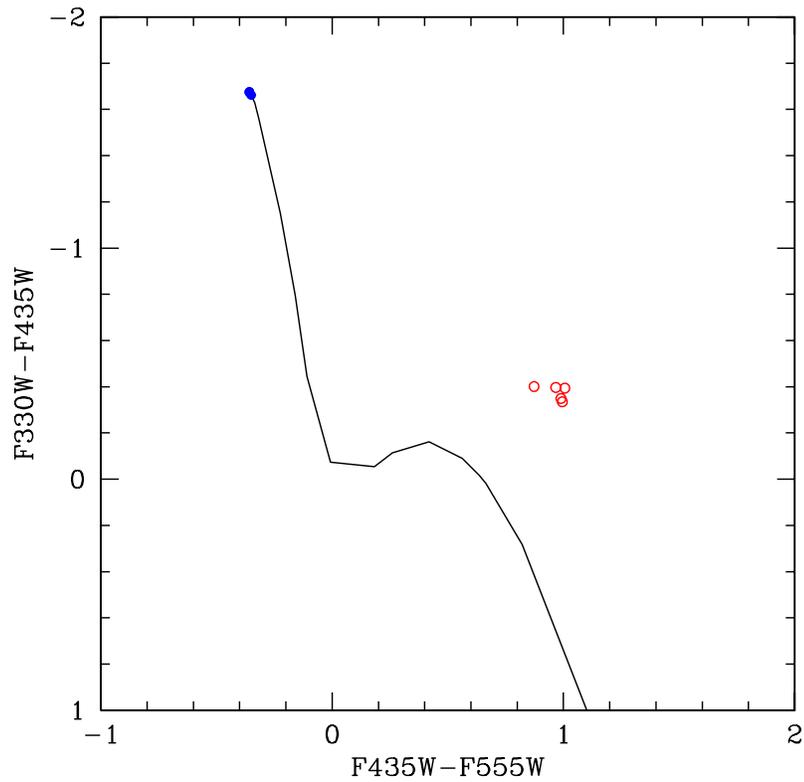}
\caption{Color-color diagram of the five MS member stars whose
spectral types are available from Melena et al.\ (2008).
 The black curve is the empirical ZAMS from Sparke \& Gallagher (2007). The red open circles denote the
observed colors of the five stars, while the blue filled circles indicate the
intrinsic colors derived from their corresponding spectral type.
 \label{Fig.10}}
\end{figure}

\begin{figure}[]
\includegraphics[width=11cm]{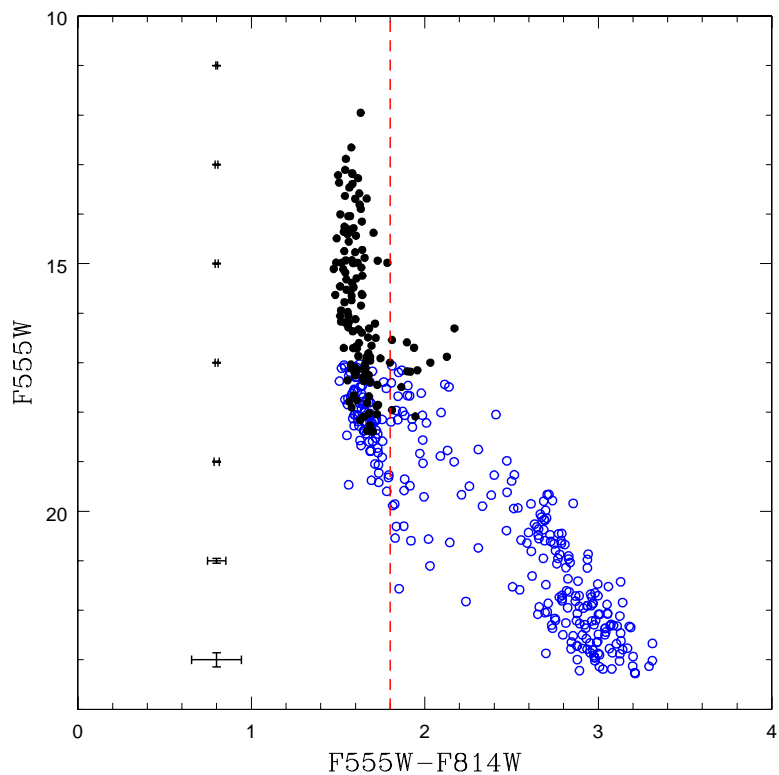}
\caption[Color-magnitude diagram ($V$ vs. $V-I$) of all member stars
on the PC chip of WFPC2] { Color-magnitude diagram of all
  proper-motion-selected member
stars with UV and optical photometry (black dots) and members
fainter than 17\,mag in {\em F555W} observed on the PC chip of the
WFPC2 (blue open circles). Stars with UV-optical photometry that are
bluer than ({\em F555W}-{\em F814W})=1.8\,mag (dashed vertical line)
are selected as probable MS stars. \label{Fig.5}}
\end{figure}

\begin{figure}[]
\includegraphics[width=11cm]{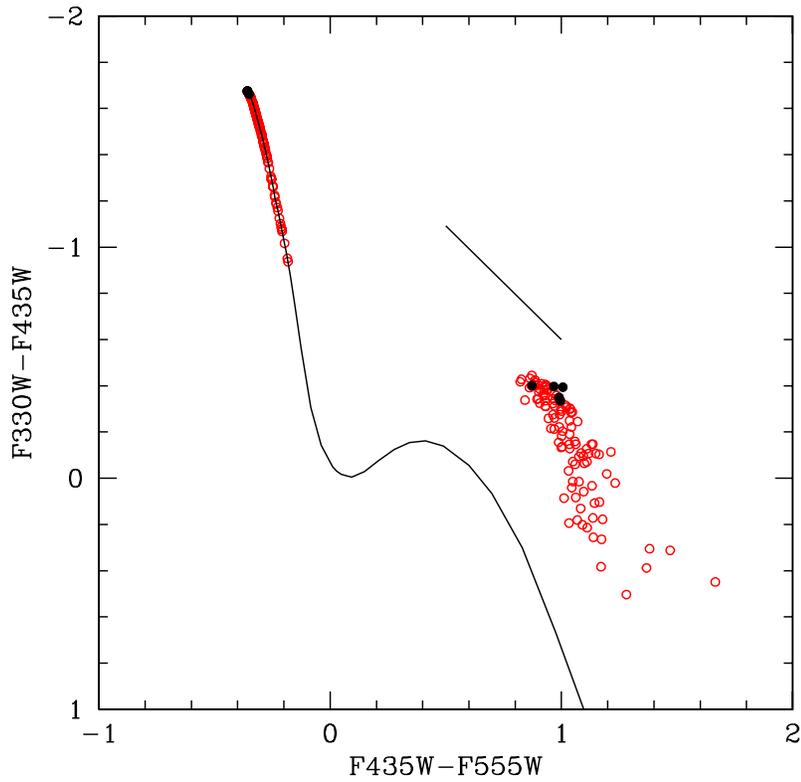}
\caption{Color-color diagram of the selected MS member stars ({\em
F555W}-{\em F814W}$\le$1.8\,mag). The black straight line indicates
the slope of the reddening vector determined from the five MS member
stars (black solid dots) whose spectral types are available from
Melena et al.\ (2008). The MS member stars (red open dots) are
projected onto the Marigo et al. (2008) isochrone (black curve) to
obtain their reddening. \label{Fig.10}}
\end{figure}

\begin{figure}[]
\includegraphics[width=\textwidth]{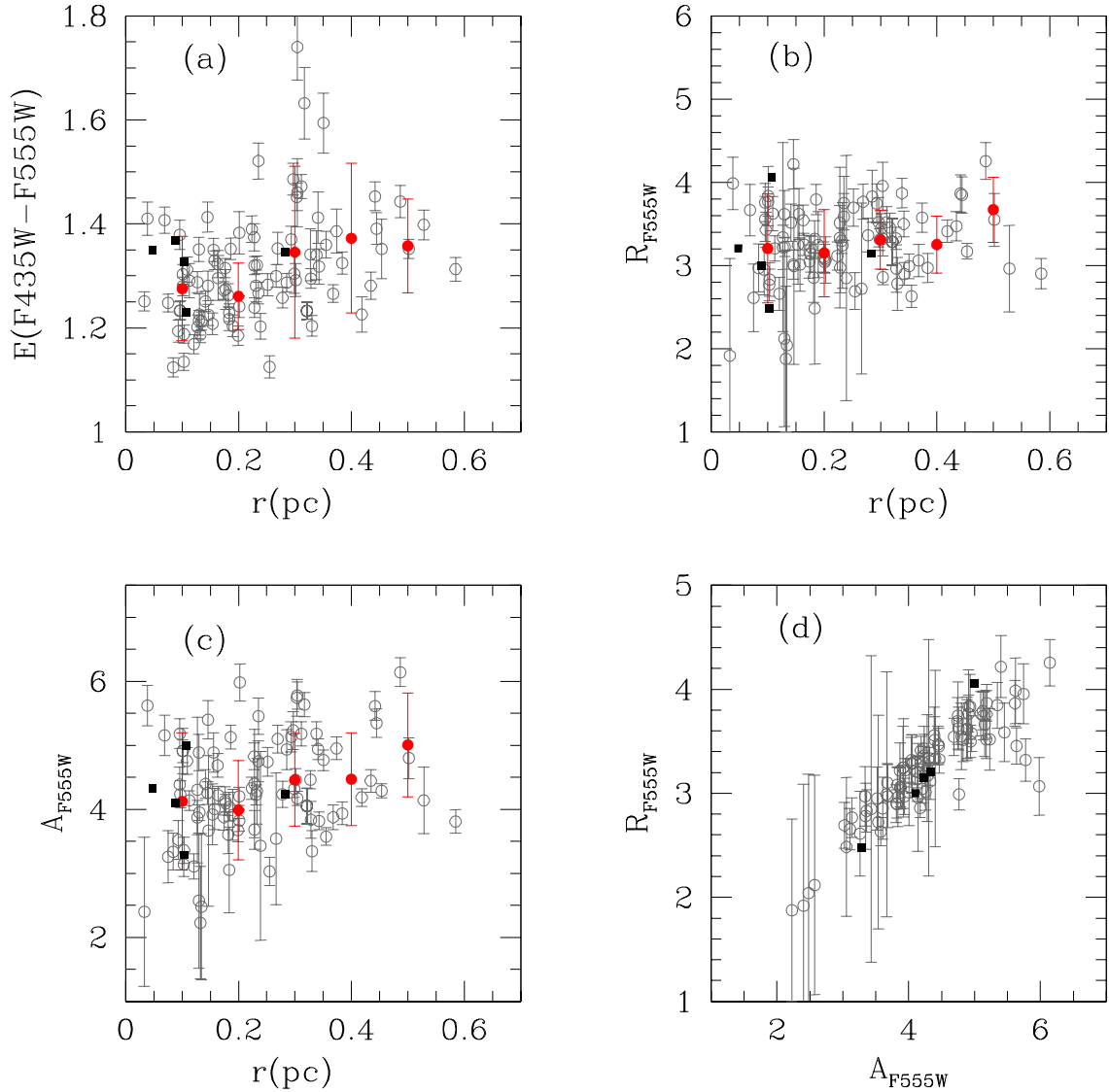}
 \caption[]{The dependence of the stellar reddening
$E(F435W-F555W)$ (a), of the $R_{F555W}$ (b), and of the $A_{F555W}$
values (c) on the cluster-centric distance without foreground
reddening correction, and the dependence of $R_{F555W}$ on
$A_{F555W}$ (d). MS members for which no spectroscopy is available
are plotted with grey open circles. The five MS member stars with
spectral information from Melena et al.\ (2008) are indicated with
filled black squares. In panel a, b, c, red filled circles and error
bars represent the mean values and dispersions of $E(F435W-F555W)$,
$R_{F555W}$ and $A_{F555W}$ respectively. They were measured in 
concentric annuli around the cluster center.  The innermost annulus was 
a circle with a radius of 0.1 pc.  The next-larger annulus ranged from 
0.1 to 0.2 pc in the inner and outer radius, respectively, and for the 
subsequent annuli the radii were increased by 0.1 pc each up to a 
maximum of 0.5 pc.
 \label{Fig.7}}
\end{figure}

\begin{figure}[]
\includegraphics[width=\textwidth]{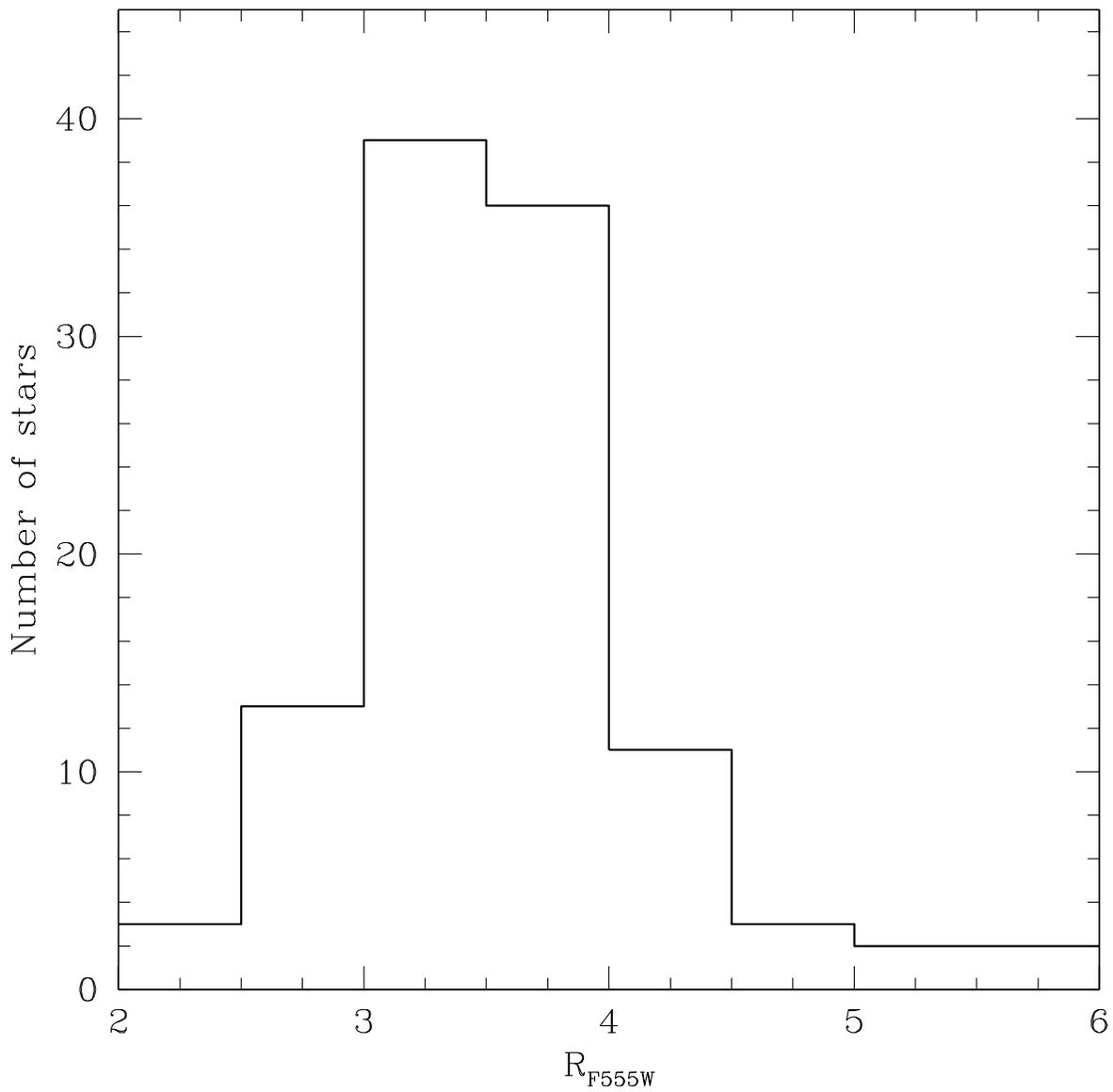}
 \caption[]{The histogram distribution of the $R_{F555W}$ values for all MS member stars ({\em
F555W}-{\em F814W}$\le$1.8\,mag).
 \label{Fig.7}}
\end{figure}

\begin{figure}[]
\includegraphics[width=\textwidth]{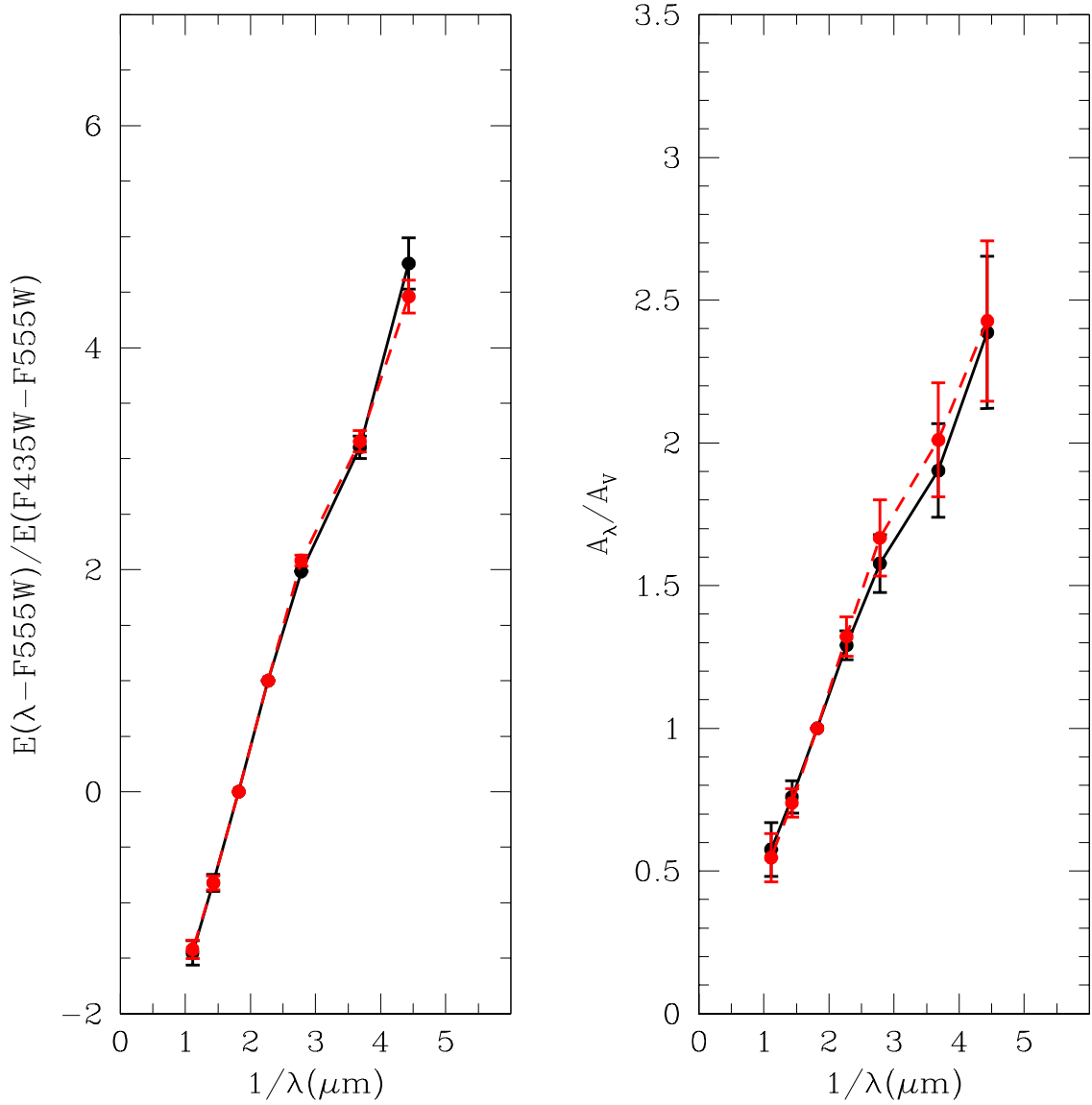}
\caption[Extinction curve for the HD\,97950 cluster] {Mean extinction
curve $E({\rm \lambda}-F555W)/E(F435W-F555W)$ (left panel) and
$A_\lambda/A_{F555W}$ (right panel) already corrected for foreground
extinction for the HD\,97950 cluster (black lines, mean extinctions of
the selected MS member stars) and the five
main-sequence member stars with spectroscopy (red dashed lines,
Melena et al. 2008) from which the cluster reddening vector is
derived (Section 3). The error bars are the standard deviation
associated with $E({\rm \lambda}-F555W)/E(F435W-F555W)$ and
$A_\lambda/A_{F555W}$ in each filter.  \label{Fig.9}}
\end{figure}

\begin{figure}[]
\includegraphics[width=\textwidth]{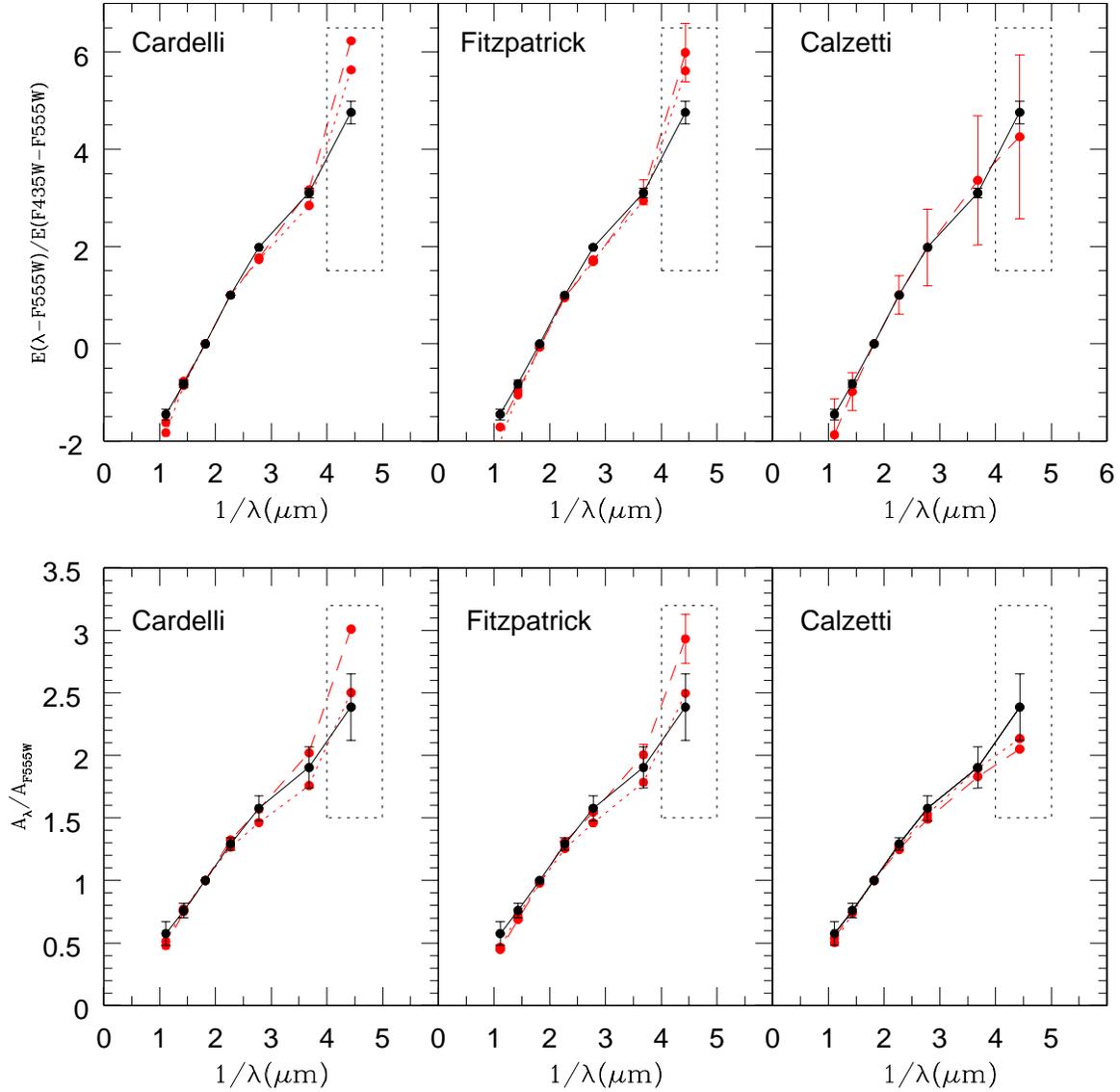}
\caption[Extinction curve for the HD\,97950 cluster] {Mean
extinction curve $E({\rm \lambda}-F555W)/E(F435W-F555W)$ (upper
panels) and $A_\lambda/A_{F555W}$ (lower panels) for the HD\,97950
cluster (black solid lines) already corrected for foreground extinction.
The red dotted lines trace Cardelli et al.'s (1989), Fitzpatrick's
(1999) and Calzetti et al.'s (2000) extinction laws with the de-reddened
value of $R_V$ as the cluster HD\,97950 ($R_V=3.75$). The red dashed
lines are the average Galactic extinction laws ($R_V=3.1$) of
Cardelli et al. (1989) and Fitzpatrick (1999), and Calzetti et al.'s
(2000) extinction law for starburst galaxies ($R_V=4.05$).
The error bars are the standard deviation associated with $E({\rm
\lambda}-F555W)/E(F435W-F555W)$ and $A_\lambda/A_{F555W}$ in each
filter. Uncertainties of Fitzpatrick's (1999) and Calzetti et al.'s
(2000) extinction laws are also indicated with error bars. For
Fitzpatrick's (1999) extinction law, its uncertainty approaches zero
for wavelengths larger than 4000\,$\rm \AA$ owing
 to the normalization to the visual band.
No such uncertainty information is available for Cardelli et al.'s (1989) extinction law.
 The dotted black boxes cover the $MUV$ filter where the
cluster extinction curve differs most dramatically from the
reference extinction laws.\label{Fig.9}}
\end{figure}

\begin{figure}[]
\includegraphics[width=\textwidth]{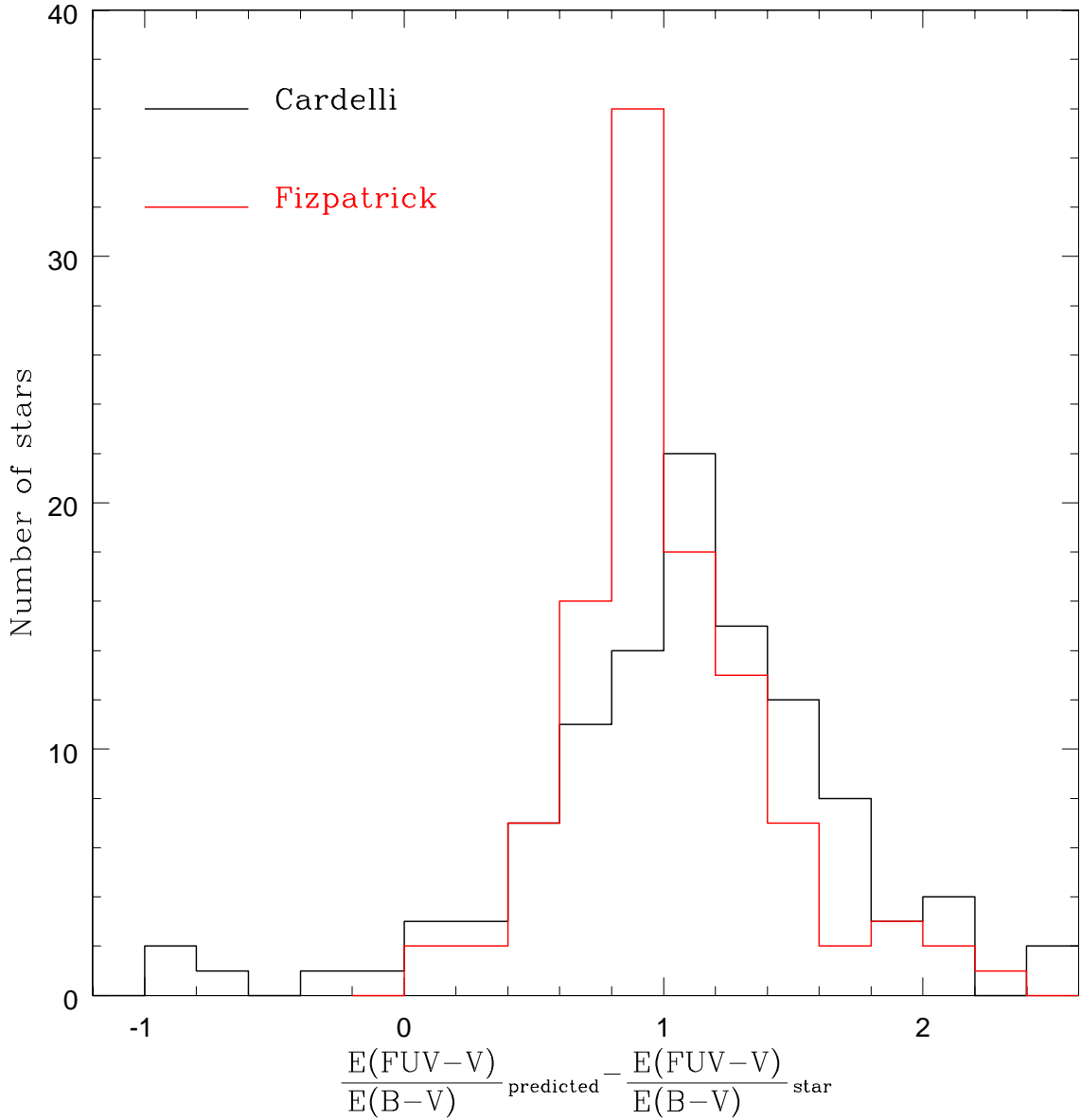}
\caption[Extinction curve for the HD\,97950 cluster] {The histogram distribution of the difference of the 
predicted reddening ratio of $E(FUV - V) / E(B - V)$ obtained for the 
Cardelli et al. (1989, black line) and the Fitzpatrick (1999, red line) 
extinction laws and the inferred extinction ratios in the corresponding 
$E(F220W - F555W) / E(F435W - F555W)$ values for the individual member 
stars in the HD 97950 cluster.  For the calculation of each difference 
the $R_{F555W}$ derived for the individual star was adopted. A
difference of positive value indicates a reduced or a missing of UV bump
for the individual member star. 
\label{Fig.9}}
\end{figure}

\begin{figure}[]
\includegraphics[width=\textwidth]{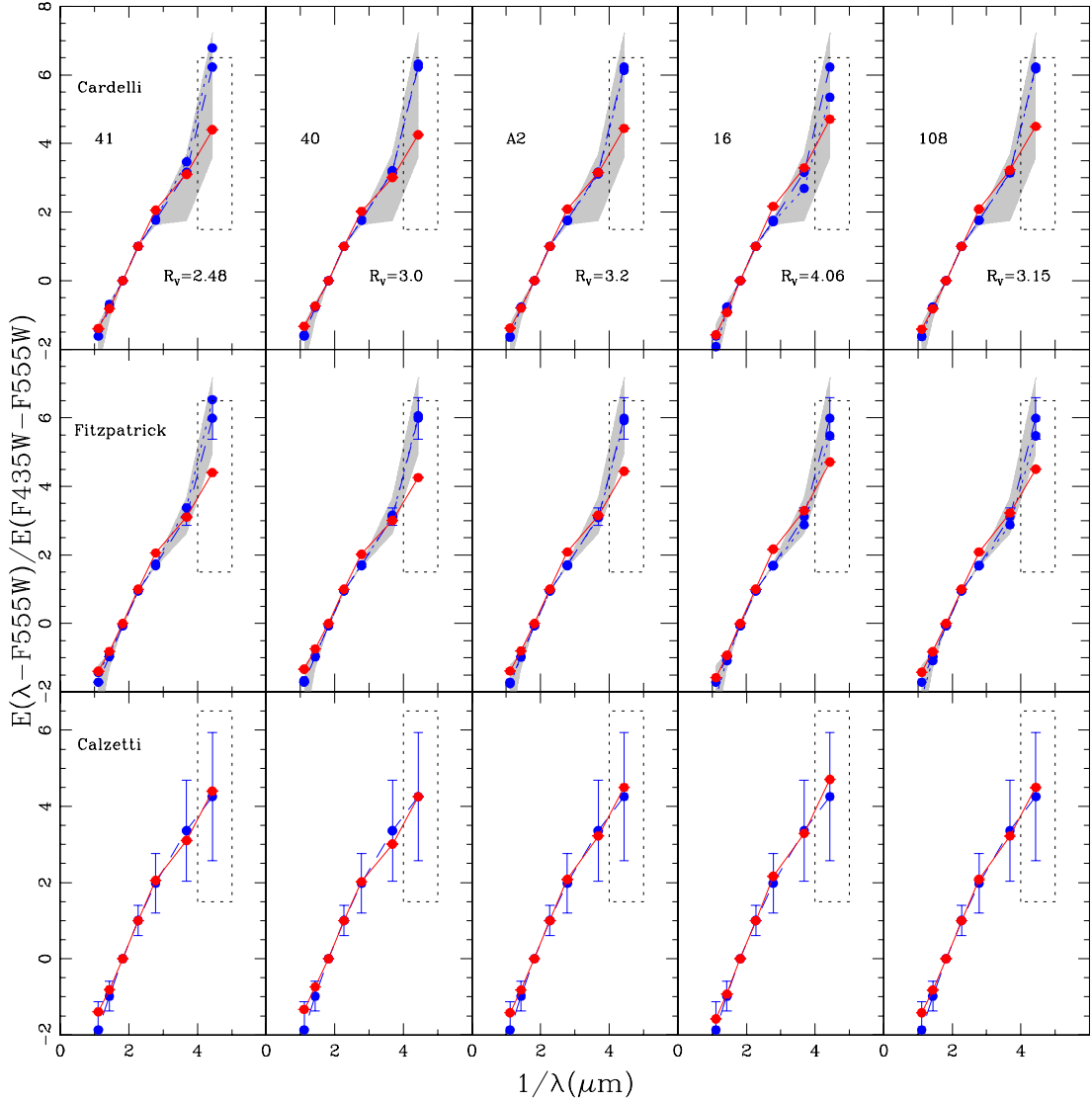}
\caption[Extinction curve for the HD\,97950 cluster] {Individual
extinction curves $E({\rm \lambda}-F555W)/E(F435W-F555W)$ for the
five MS member stars with available spectroscopy from Melena et al.\
(2008) (red lines) already corrected for foreground extinction
(panels of the same column). The error bar of
$E({\rm \lambda}-F555W)/E(F435W-F555W)$ in each filter is of the size
of data points. The designations of the stars as given by Melena et
al. (2008) are indicated in the upper panels. The blue dashed lines
(panels of the same row) are the average Galactic extinction laws
($R_V=3.1$) of Cardelli et al. (1989) and Fitzpatrick (1999), and
Calzetti et al.'s (2000) extinction law for starburst galaxies
($R_V=4.05$). The blue dotted lines trace Cardelli et al.'s (1989)
and Fitzpatrick's (1999) laws with the same value of $R_V$ as for
each individual star (indicated in the upper panels). The grey
shaded regions are Cardelli et al.'s (1989) and Fitzpatrick's (1999)
laws with $R_V$ between 2.0 (upper boundary) and 6.0 (lower
boundary).
 The meaning of the error bars and the dotted black box are the
same as in Figure 9. \label{Fig.9}}
\end{figure}


\begin{landscape}

\begin{table}\footnotesize
\centering \caption{The UV and optical photometry of member stars in
the HD\,97950 cluster} \label{tab:jj:def}
\begin{tabular}{ccccccccccccccccc}
\hline
No. & $R.A.$ & $Dec$ & {\em F220W} & $\sigma_{220}$ & {\em F250W} & $\sigma_{250}$ & {\em F330W} & $\sigma_{330}$ & {\em F435W} & $\sigma_{435}$ & {\em F555W} & $\sigma_{555}$ & {\em F675W} & $\sigma_{675}$ & {\em F814W} & $\sigma_{814}$ \\
\hline
    1 & 168.7816010 & -61.2605972 & 15.34 & 0.003 & 13.51 & 0.002 & 12.64 & 0.002 & 12.99 & 0.002 & 11.95 & 0.003 & 11.00 & 0.003 & 10.32 & 0.023   \\
   2 & 168.7804565 & -61.2608185 & 15.86 & 0.003 & 14.05 & 0.002 & 13.23 & 0.002 & 13.62 & 0.002 & 12.65 & 0.005 & 11.74 & 0.004 & 11.08 & 0.027   \\
   3 & 168.7784424 & -61.2599373 & 16.00 & 0.003 & 14.24 & 0.003 & 13.42 & 0.003 & 13.85 & 0.003 & 12.89 & 0.005 & 12.00 & 0.005 & 11.34 & 0.030   \\
   4 & 168.7788391 & -61.2608719 & 16.13 & 0.004 & 14.39 & 0.003 & 13.62 & 0.003 & 14.06 & 0.003 & 13.19 & 0.006 & 12.26 & 0.005 & 11.61 & 0.030   \\
   5 & 168.7794037 & -61.2609558 & 16.26 & 0.004 & 14.48 & 0.003 & 13.66 & 0.003 & 14.08 & 0.003 & 13.11 & 0.005 & 12.22 & 0.005 & 11.57 & 0.028   \\
   6 & 168.7803192 & -61.2597466 & 16.49 & 0.004 & 14.64 & 0.003 & 13.77 & 0.003 & 14.15 & 0.003 & 13.18 & 0.005 & 12.25 & 0.005 & 11.59 & 0.028   \\
   7 & 168.7788544 & -61.2609329 & 16.40 & 0.004 & 14.65 & 0.003 & 13.86 & 0.003 & 14.28 & 0.003 & 13.39 & 0.006 & 12.45 & 0.005 & 11.81 & 0.028   \\
   8 & 168.7759705 & -61.2602043 & 16.16 & 0.004 & 14.46 & 0.003 & 13.70 & 0.003 & 14.15 & 0.003 & 13.21 & 0.006 & 12.34 & 0.005 & 11.71 & 0.042   \\
   9 & 168.7806244 & -61.2607231 & 16.48 & 0.004 & 14.68 & 0.004 & 13.82 & 0.004 & 14.20 & 0.003 & 13.28 & 0.006 & 12.36 & 0.007 & 11.66 & 0.040   \\
  10 & 168.7788086 & -61.2602081 & 16.35 & 0.004 & 14.60 & 0.003 & 13.83 & 0.003 & 14.25 & 0.003 & 13.36 & 0.006 & 12.49 & 0.005 & 11.86 & 0.031   \\
\hline
\end{tabular}
\begin{flushleft}
\indent Note.----Table 1 is presented in its entirety in the electronic edition of the Astronomical Journal.
A portion is shown here for guidance regarding its form and content.\\
\indent $R.A.$ and $Dec$ are Right Ascension and Declination of the
member stars on the PC chip of WFPC2 based on the 2007 data. They
are J2000 coordinates. The {\em F220W}, {\em F250W}, {\em F330W},
and {\em F435W} photometry is from ACS.  The {\em F555W}, {\em
F675W}, and {\em F814W} photometry is from WFPC2 (2007 data). All
magnitudes are Vega magnitudes in the ACS/HRC and WFPC2 filter
systems.

\end{flushleft}
\end{table}

\begin{table}\footnotesize
\centering \caption{Photometry of main sequence stars with spectral
types} \label{tab:jj:def}
\begin{tabular}{cccccccccccccccccc}
\hline Designation & $R.A.$ & $Dec$ & {\em F220W} & {\em F250W} & {\em
F330W} & {\em F435W} & {\em F555W} & {\em F675W} & {\em F814W} &
Spectral
type & $(U-B)_0$ & $(B-V)_0$ \\
\hline
 41 & 168.7793427 & -61.2608604 & 17.50 & 15.70 & 14.91 & 15.31 & 14.44 & 13.50 & 12.83 & O4V & -1.210 & -0.320 \\
40 & 168.7796783 & -61.2608986 & 16.69 & 14.88 & 14.07 & 14.47 & 13.46 & 12.59 & 11.90 & O3V  & -1.221 & -0.320\\
A2 & 168.7804565 & -61.2608185 &  15.86 & 14.05 & 13.23 & 13.62 & 12.66 & 11.74 & 11.08 & O3V & -1.221 & -0.320 \\
16 & 168.7825775 & -61.2605515 & 17.03 & 15.20 & 14.33 & 14.68 & 13.69 & 12.77 & 12.09 & O3V  & -1.221 & -0.320\\
108 & 168.7854614 & -61.2606812 & 17.33 & 15.50 & 14.56 & 14.89 & 13.90 & 12.96 & 12.27 & O5.5V & -1.183 & -0.321\\
\hline
\end{tabular}
\begin{flushleft}
\indent Note.----All coordinates are J2000 coordinates. The colors
$(U-B)_0$ and $(B-V)_0$ are taken from the empirical ZAMS of Sparke
\& Gallagher (2007).
\end{flushleft}
\end{table}

\begin{table}\footnotesize
\centering \caption{The color excesses and extinctions of member
stars} \label{tab:jj:def}
\begin{tabular}{ccccccccccccccccc}
\hline No. & $R.A.$ & $Dec$ & {\tiny \em E(MUV-B)} & {\tiny \em E(NUV-B)}
& {\tiny \em E(U-B)} & {\tiny \em E(B-V)} & {\tiny \em E(V-R)} &
{\tiny \em E(V-I)} & $\tiny A_{F220W}$ & $A_{F250W}$
& $A_{F330W}$ & $A_{F435W}$ & $A_{F555W}$ & $A_{F675W}$ & $A_{F814W}$ & $R_{F555W}$ \\
\hline
  1 & 168.7788391 & -61.2608719 &  4.76 & 2.599 &  1.17 & 1.194 &  1.12 & 1.920 &   8.42 &  6.259 &  4.84 & 3.660 &  2.47 & 1.347 &  0.55 & 2.065      \\
  2 & 168.7788544 & -61.2609329 &  4.82 & 2.639 &  1.20 & 1.220 &  1.12 & 1.920 &   8.86 &  6.677 &  5.24 & 4.038 &  2.82 & 1.694 &  0.90 & 2.310      \\
  3 & 168.7806244 & -61.2607231 &  4.98 & 2.751 &  1.24 & 1.258 &  1.11 & 1.953 &   9.06 &  6.830 &  5.32 & 4.079 &  2.82 & 1.715 &  0.87 & 2.242      \\
  4 & 168.7788086 & -61.2602081 &  4.81 & 2.636 &  1.20 & 1.222 &  1.06 & 1.844 &   9.24 &  7.068 &  5.63 & 4.432 &  3.21 & 2.149 &  1.37 & 2.627      \\
  5 & 168.7812195 & -61.2629204 &  5.10 & 2.825 &  1.30 & 1.323 &  1.17 & 2.016 &  11.80 &  9.528 &  8.00 & 6.703 &  5.38 & 4.209 &  3.36 & 4.067      \\
  6 & 168.7785034 & -61.2624550 &  5.01 & 2.776 &  1.26 & 1.284 &  1.12 & 1.968 &   9.80 &  7.562 &  6.05 & 4.786 &  3.50 & 2.385 &  1.53 & 2.727      \\
  7 & 168.7825775 & -61.2605515 &  5.17 & 2.879 &  1.33 & 1.347 &  1.11 & 1.952 &  12.37 & 10.080 &  8.53 & 7.201 &  5.85 & 4.742 &  3.90 & 4.346      \\
  8 & 168.7780762 & -61.2599411 &  4.83 & 2.670 &  1.22 & 1.235 &  1.09 & 1.895 &   9.31 &  7.146 &  5.69 & 4.476 &  3.24 & 2.156 &  1.35 & 2.624      \\
  9 & 168.7793427 & -61.2608604 &  4.84 & 2.619 &  1.17 & 1.194 &  1.13 & 1.939 &   9.47 &  7.250 &  5.81 & 4.631 &  3.44 & 2.310 &  1.50 & 2.879      \\
 10 & 168.7854614 & -61.2606812 &  5.20 & 2.936 &  1.32 & 1.342 &  1.12 & 1.973 &  10.84 &  8.578 &  6.96 & 5.642 &  4.30 & 3.176 &  2.33 & 3.204      \\
\hline
\end{tabular}
\begin{flushleft}
\indent Note.----Table 3 is presented in its entirety in the
electronic edition of the Astronomical Journal.
A portion is shown here for guidance regarding its form and content.\\
\indent $RA$ and $DEC$ are Right Ascension and Declination of the
member stars on the PC chip of WFPC2 based on the 2007 data. They
are J2000 coordinates. For convenience, the color excess
$E(F220W-F435W)$, $E(F250W-F435W)$, $E(F330W-F435W)$,
$E(F435W-F555W)$, $E(F555W-F675W)$, and $E(F555W-F814W)$ are named
$E(MUV-B)$, $E(NUV-B)$, $E(U-B)$, $E(B-V)$, $E(V-R)$, and $E(V-I)$,
respectively.

\end{flushleft}
\end{table}

\end{landscape}

\begin{deluxetable}{cr}[]
\tablecolumns{8} \tablewidth{0pc} \tablecaption{Mean color excesses and extinctions for MS member stars in the cluster
 (without foreground reddening correction)}
\tablehead{ \colhead{Color excess \& extinction}
& \colhead{Values} }
\startdata
  $E(F220W-F435W)$ &  $4.97\pm0.21$  \\
  $E(F250W-F435W)$ &  $2.78\pm0.18$  \\
  $E(F330W-F435W)$  &   $1.31\pm0.12$  \\
  $E(F435W-F555W)$  &  $1.33\pm0.12$   \\
  $E(F555W-F675W)$  &  $1.08\pm0.05$   \\
  $E(F555W-F814W)$  &  $1.91\pm0.06$   \\
$A_{F220W}$ & $11.01\pm0.12$ \\
$A_{F250W}$ & $8.83\pm0.12$ \\
$A_{F330W}$ & $7.35\pm0.12$    \\
$A_{F435W}$ & $6.05\pm0.10$    \\
$A_{F555W}$ & $4.72\pm0.10$    \\
$A_{F675W}$ & $3.64\pm0.10$    \\
$A_{F814W}$ & $2.80\pm0.10$    \\
$R_{F555W}$ &  $3.54\pm0.63$   \\
\enddata
\end{deluxetable}

\end{document}